# THEORY OF ACOUSTIC EMISSION FOR MICRO-CRACKS APPEARED UNDER THE SURFACE LAYER MACHINING BY COMPRESSED ABRASIVE

A.K. Aringazin,[1,3] V.D. Krevchik,[2,3]
V.A. Skryabin,[2] M.B. Semenov,[2,3] G.V. Tarabrin[2]

[1] Eurasian National University, Astana 010008, Kazakhstan
aringazin@mail.kz

[2] Physics Department, Penza State University, Penza 440017, Russia
physics@diamond.stup.ac.ru

[3] Institute for Basic Research, P.O. Box 1577, Palm Harbor, FL 34682, USA
ibr@verizon.net

One of the possible mechanisms for acoustic emission of growing micro-cracks under conditions of the material machining by compressed abrasive has been theoretically studied. Physical ground of this mechanism is the dislocation creep in the field of instant contact temperature on stage of micro-cutting with appearance of the wedge-shaped cavity. It has been shown that the energy density for radiated acoustic wave at the moment when the cavity is opened essentially depends on parameters of the material abrasive machining.

## 1. Introduction

Interest to acoustic emission for micro-cracks is stimulated by possibilities to use it in nondestructive control of mechanical state of material in processes of its machining or exploitation [1-3]. Also the interest is due to fundamental aspect which is related to application of the acoustic emission method to research micro-cracks growth physical mechanisms [2, 4]. Some general method of analysis of the acoustic emission of arbitrary micro-cracks in the restricted elastic solids has been developed in Ref. [1]. This method is based on the Huygens principle and allows one to determine both the law of motion of micro-cracks edges under influence of applied external stress and the spectral density of acoustic emission.

However, theoretical approach developed in Ref. [1] does not contain any real physical mechanisms of the micro-cracks appearance prior to its opening. Investigations of these mechanisms are especially important to understand processes of machining of material by a compressed abrasive. Indeed, heat processes which are initiated by instant contact temperature on the stage of micro-cutting can stimulate structural changes of the defects in material surface layer. These changes are related to motion of dislocations. Besides, accumulation of dislocations near to impurity is one of the possible mechanisms of the formation of micro-cracks [4].

## 2. The theoretical model of acoustic emission

Mechanism of acoustic emission for micro-cracks, which are appeared under process of dislocation strengthening during micro-cutting in the material surface layer of the machining sample, has been theoretically investigated in this paper within the framework Huygens principle for elastic material. Physical grounds of this mechanism are represented by the dislocation creep, which is stimulated by temperature pulse. As the result of this, a number of parallel dislocations accumulate near to obstacle, which is appeared as impurity release. In the region of such accumulation, the wedge-shaped cavity can arise, and this cavity can be a source of the micro-crack appearance. Acoustic emission takes place in the moment of the cavity opening; and, as it will be demonstrated, its spectral density of radiation essentially depends from parameters of dislocation creep and from the material abrasive machining.

As it is known [5] in the process of micro-cutting, strong temperature pulses can arise. Characteristic time for this process $t_0$ is determined by the mean width of micro-ledges $\overline{L}_m$ and the mean velocity $\overline{\upsilon}$ of the abrasive grain: $t_0 = \overline{L}_m / \overline{\upsilon}$. For example, it equals $t_0 = 10^{-5} s$ for $\overline{\upsilon} = 2\, m/s$ and $\overline{L}_m = 200\, mcm$. As the result, dislocations can be released from fixed impurity and under any critical stress dislocations can drift together with an impurity cloud [4]. Estimation of the instant contact temperature can be made in terms of simplest boundary-value problem with the instant heat source $F(x,t) = Q_0 \delta(x-\xi)\delta(t-\tau)$, where $Q_0 = 2M\overline{\upsilon}^2 / (\pi \overline{L}^2)$ is the source intensity, $M$ is the mean value of the abrasive grain mass.

It is then easy to demonstrate that value of the instant contact temperature can be represented as:

$$\langle T(x,t) \rangle_e = \frac{Q_0}{cg} \cdot \frac{1}{2\sqrt{\pi}} \cdot \frac{\Lambda}{L^2} Arcth \sqrt{\frac{2a\sqrt{t}}{L}+1}, \qquad (1)$$

where the brackets in the left-hand-side of Eq. (1) mean that $T(x,t)$ is meant with respect to the dislocation effective length $\ell$ with the weight function $N(\ell)$ of the Gauss type $N(\ell) = \Lambda \exp(-\ell^2/L^2)/(\sqrt{\pi}L^2)$, where $\Lambda$ is the common dislocation length in the part material, $L$ is the mean length of dislocation loop, $c$ is the material specific heat, $\rho$ is the material density, $a^2$ is the temperature transport coefficient.

Let us estimate $\langle T(x,t) \rangle_e$ for the following numerical values: $\Lambda = 6 \cdot 10^{-2}\, m, L = 60\, mcm, c = 460\, J/(kg \cdot K), \rho = 7800\, kg/m^3, Q_0 \approx 10^3\, J/m^2$, $M = 6{,}7\, m^2$ for the parameter of granularity $80, t = t_0 = 10^{-5} s$. The estimation is $\langle T(x,t) \rangle_e \approx 1820\, K$. One can se that the instant contact temperature can be as high as the metal melting temperature [5]. Dislocation can escape the impurity cloud, with subsequent joint drift in the temperature field, under any critical stress, which can arise, for example, during micro-cutting. Under these conditions, the critical drift speed $\upsilon_ê$ can be determined as [4]:

$$\upsilon_ê = \frac{8D(t)}{2r_0}, \qquad (2)$$

where $D(t) = D_0 \exp\left[-Q/\left(k\langle T(x,t)\rangle_e\right)\right]$ is the diffusion coefficient for fixing impurity in field of the instant contact temperature; $Q$ is the diffusion activation energy; $r_0$ is the radius of Kotrell cloud around dislocation loop.

Taking into account Eq. (2), the critical speed of creep $\dot{\varepsilon}_ê$ can be written as [4]:

$$\dot{\varepsilon}_ê = \frac{8D(t)}{2r_0}bN_d, \qquad (3)$$

where $b$ is the Burgers' vector value; $N_d$ is the dislocation density in material of machining part.

As it has been mentioned above, during the drift some part of dislocations is joined together and can form the wedge-shaped cavity (near obstacle) [4]. In the region of such accumulation, a stable cavity of length $L_n = n^2 b$ can arise from $n$ dislocations that can lead to appearance of micro-cracks. Substituting $n \approx \dot{\varepsilon}_ê \tau$, where $\tau = 2\pi/\omega$ is the time period up to the micro-cracks growth begins and $\omega$ is the angular speed for the machining part rotation, we obtain for $L_n$:

$$L_n = b\left[\frac{8bN_d}{2r_0}D(t)\tau\right]^2. \qquad (4)$$

As we can see from Eq. (4), the growth value for the wedge-shaped cavity is determined mostly by the diffusion mobility of impurity, which "decorates" dislocation, and also by the dislocation density and of course by the parameters of the material abrasive machining,. $\bar{\upsilon}, \omega, u, M$. Calculation of $L_n$ for the following values:
$b = 10^{-10} m, \omega = 5\, rad/s, D_0 \approx 10^{-7} m^2/s, Q \approx 1,3\, eV, N_d = 10^{16}\, m^{-2}, t_0 = 10^{-5} s,$
$D(t_0) \approx 2,8 \cdot 10^{-11} m^2/s, 2r_0 = 0,1\, \mu m, L \approx 60\, \mu m,$ gives $L_n \approx 0,85\, mm$.

Theoretical approach to the cavity acoustics is based on the Huygens principle for solid elastic media [1] with account of dislocation creep in conditions of the material abrasive machinery. Since the cavity is a primary medium cut, which is opened by influence of applied mechanical stresses without change of effective length $L_n$ such a situation is equivalent to model of the "instantly spreading micro-crack" [1]. Here, the energy density $W$ for radiating acoustic wave, as it has been demonstrated earlier [4], is approximately determined as

$$W \approx \frac{\sigma_0^2 L_n^4 \omega_s^4}{64\pi^6 \rho c_e^6}, \qquad (5)$$

where $\sigma_0$ is the amplitude of mechanical stress; $\omega_s$ is the frequency of the acoustic wave; $c_e$ is the limit speed for lowest symmetric Lamb mode [1].

Substituting $\sigma_0 = 10^5 \, N/m^2$, $L_n = 0,85 \, mm$, $\omega_s = 3 \, MHz$, we obtain from Eq. (5) $W \approx 82 \, \mu J/m^3$, amd the radiation intensity is $I \approx 0,05 \, mW/cm^2$.

## 3. Conclusions

One of the possible mechanisms of the micro-cracks appearance, which is related to dislocation creep initiated by field of the instant contact temperature on the stage of micro-cutting, has been developed.

Our numerical estimations show that the surface material layer, as the result of the micro-cutting process, gets dislocation strengthening, under which the wedge-shaped cavity can arise. Also, the acoustic emission takes place at the moment when the cavity is opened. This can be identified experimentally.